# Study the effect of scratching depth and ceramic-metal ratio on the scratch behavior of NbC/Nb Ceramic/Metal nano-laminates using molecular dynamics simulation and machine learning

**Mesbah Uddin[1], Iman Salehinia[2]**
Department of Mechanical Engineering, Northern Illinois University[1,2]
590 Garden Road, DeKalb, IL, 60115.

**ABSTRACT**

Developing a new class of coating materials is necessary to meet the increasing demands of energy and defense-related technologies, aerospace engineering, and harsh environmental conditions. Functional-based coatings, such as ceramic-metal nanolaminates, have gained popularity due to their ability to be customized according to specific requirements. To design and develop advanced coatings with the necessary functionalities, it is crucial to understand the effects of various parameters on the mechanical and tribological properties of these coatings. In this study, we investigate the impact of penetration depth, individual layer thickness, and ceramic-metal ratio on the mechanical and tribological properties of ceramic-metal nanolaminates, particularly NbC/Nb. Our findings reveal that the thickness of the individual metallic and ceramic layers significantly affects the coatings' properties. However, some models exhibited punctures on the top ceramic layer, which altered the scratching behavior and reduced the impact of layer thickness on it. This is because the top ceramic layer's thickness is too low, and the indenter can easily puncture it instead of pushing the ceramic atoms. The minimum thickness required to resist indentation is called the critical thickness, which depends on the indentation size and penetration depth. In the latter part of this paper, we employed machine learning to reduce computational costs, and the model predicts the friction coefficient with an R-squared value of 0.958.

---

[1] Corresponding author, Email: mesbahuddin1991@gmail.com

1. **INTRODUCTION**

A functionally based coating material is a type of composite material that has varying compositions in different areas depending on the performance requirements. Ceramic-metal nanolaminates are a promising option for this type of coating due to their unique properties. By changing the combination of different geometric parameters of ceramic-metal nanolaminates, we can create coatings with different functionalities. These materials offer exceptional properties such as high strength, hardness, and wear resistance [1-4]. They can also be used as advanced materials for various applications, including engine parts, medical implants, the automotive industry, aerospace, and defense industry. Scratching of CMLs can occur due to external forces acting on the material, such as during manufacturing or application, or due to wear and tear during service. The scratching process involves the interaction of the tip and the surface, resulting in deformation and material removal. Scratching can lead to various surface defects, including cracks, delamination, and wear. Understanding the scratching behavior of CMNs is important for improving their performance in practical applications. In recent years, computation tools have been used for different studies [5-15], i.e., molecular dynamics (MD) simulations have emerged as a powerful tool for investigating the scratching behavior of CMNs at the atomic scale [16-40].

The scratching behavior of CMLs is influenced by several factors. The effect of those various factors such as interface type [41, 45], scratching speed [45, 46, 47, 48], scratching direction [45, 49], temperature [45, 50], indenter size [51, 52], shape [45, 53], scratching depth [41, 49, 52, 54, 55] on the mechanical and tribological behavior have been investigated. These studies have reported a signification effect of these parameters on the scratching behavior. The current study on scratching behavior of ceramic/metal multilayer materials at the atomistic size scale has received limited attention, particularly with regards to the effect of scratching depth as most of the previous studies are done in nanoscale. Understanding the impact of scratching depth on the mechanical response of these materials is crucial for realistic interpretation of computational results in materials design, particularly in applications beyond the nanoscale. However, the high computational cost associated with molecular dynamics atomistic simulations often causes researchers to overlook the effects of penetration depth, leading to an incomplete understanding of the mechanical behavior of materials at different scales. Furthermore, the impact of design parameters, such as individual layer thickness, on scratching behavior has not been thoroughly explored. Given that both individual layer thickness and ceramic-metal ratio significantly influence the mechanical behavior of nanolaminates, it is crucial to control these factors during fabrication for tribological applications. In a previous study, we investigated the impact of individual layer thickness on the scratching behavior of NbC/Nb multilayers. However, the scratching only penetrated 3nm, indicating that only the first few layers of ceramic and a small section of the metallic layer were affected. Therefore, this study did not provide a comprehensive understanding of the effects of individual layer thickness.

This academic paper builds upon our prior research and employs molecular dynamics atomistic simulations to investigate how scratch depth impacts the scratching behavior of NbC/Nb multilayer samples with different metal thicknesses. In addition, the individual layer thickness dependency on the scratching behavior was captured.

In addition, molecular dynamics simulation cost is very expensive and time consuming. Recently machine learning has become a popular method to predict different features [63]. In the last section of this paper, a machine-learning approach has been used to predict the friction coefficient of different models of ceramic-metal nanolaminate.

2. MODELING

LAMMPS software was utilized to conduct molecular dynamics simulations with the second nearest neighbor modified embedded atom method (2NN-MEAM) interatomic potentials for NbC and Nb. These potentials have been previously employed by Salehinia et al. to examine the deformation mechanisms of NbC/Nb under uniform compression and nano-indentation. Additionally, the potential was able to accurately reproduce various physical properties of metals and ceramics, including lattice parameters, enthalpy of formation, elastic properties, surface energies, and interface energies. The interface of NbC/Nb was found to adopt the Baker-Nutting orientation relationship (OR), where the (001) planes of Nb and NbC single crystals form the interface, and the [100] crystallographic direction of NbC is parallel to the [110] direction of Nb [59].

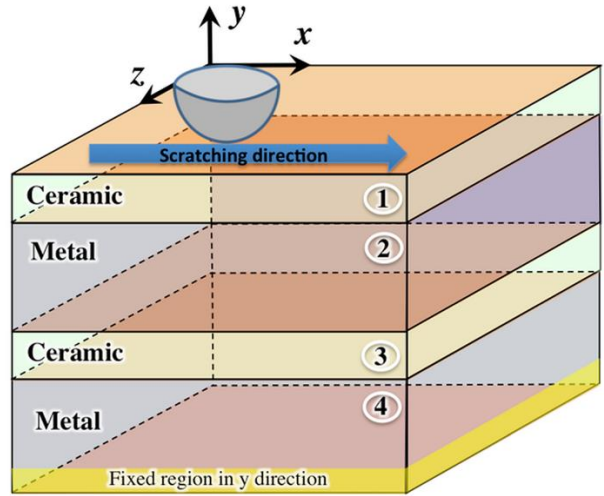

Fig. 1- The simulation cell used for nanoindentation and scratching. The layers are numbered 1-4 from top to bottom. Only the thickness of layer 2 was varied.

The utilized multilayer samples in this study consisting of alternating layers of ceramic and metal, with a minimum of four layers and a fixed 2 nm thickness for the ceramic layers and 6 nm thickness for the bottom metal layer to replicate the behavior of ceramic/metal nanolaminates, with the bottom metal layer acting as an elastic foundation for the layers above. The width of each multilayer was chosen to minimize coherency strains and boundary effects, and periodic boundary conditions were applied to the side faces. The thickness of the first metallic layer was varied to investigate its effect on the mechanical and tribological behavior of the samples. The top ceramic and metal layer thickness were varied and used in this study.

Table 1- The details of the considered models.

|  | Layer 1 | Layer 2 | Layer 3 | Layer 4 | Number of atoms |
|---|---|---|---|---|---|
| Material | Ceramic | Metal | Ceramic | Metal |  |
| CMNLI, 2226 (22) | 2 nm | 2 nm | 2 nm | 6 nm | 648584 |

| CMNLII, 2826 (28) | 2 nm | 8 nm | 2 nm | 6 nm | 914840 |
|---|---|---|---|---|---|
| Nb single crystal | Model thickness is 12 nm | | | | 657320 |
| NbC single crystal | Model thickness is 12 nm | | | | 820864 |

The models were subjected to periodic boundary conditions on their side faces, with the top surface left unrestricted and the bottom surface fixed to prevent any rigid movements of the model while undergoing indentation. A rigid spherical nano-indenter was used with a set penetration depth of 7 nm to ensure adequate penetration of the metallic layers. One should note that for 7 nm penetration depth, the indenter radius cannot be 5 nm, therefore, the indenter radius was kept at 10 nm. The indenter speed was set to 100 m/s for nano-indentation and 250 m/s for nano-scratching, with a scratching length of 20 nm. Prior to equilibration, an energy minimization using conjugate gradient method was applied. Dynamic relaxation was then used to equilibrate the models at 10 K and 0 pressure using the NPT (isothermal-isobaric) ensemble for 10 ps. The simulation used a microcanonical ensemble during nano-indentation and nano-scratching, while temperature was maintained at 10 K to better observe deformation mechanisms.

## 3. RESULTS AND DISCUSSION

The effect of penetration depth on the scratching behavior of NbC/Nb laminates was discussed here.

Figure 1 shows the removed material volume for different models. It is shown that both CMNLs models produce a high volume of removed material at higher penetration depth. This is in contrast to our study at 3 nm penetration depth where the simulations only showed material removal for the model with the thinnest metallic layers, i.e., 22. In addition, the CMNL model with thicker metallic layers displayed the highest removed material volume at 7 nm penetration depth, followed by the CMNLI and NbC single crystal models. To comprehend this behavior, it is necessary to examine the atomic snapshot during scratching. Figure 14 presents the perspective and front views of the scratched surface at the end of a 20 nm scratching length for both models at 7 nm penetration depth. For ease of visualization, all layers except the top ceramic layer have been removed. The images demonstrate that the indenter punctured the top ceramic layer for both models, which is caused by the larger attack angle, defined as the acute angle between the tangent of the semi-sphere indenter at the contact radius and the model's top surface, increasing with penetration depth. Once the ceramic layer is punctured, atoms easily pile up around the indenter, regardless of the metal-ceramic ratio, resulting in a large amount of material removal in both models.

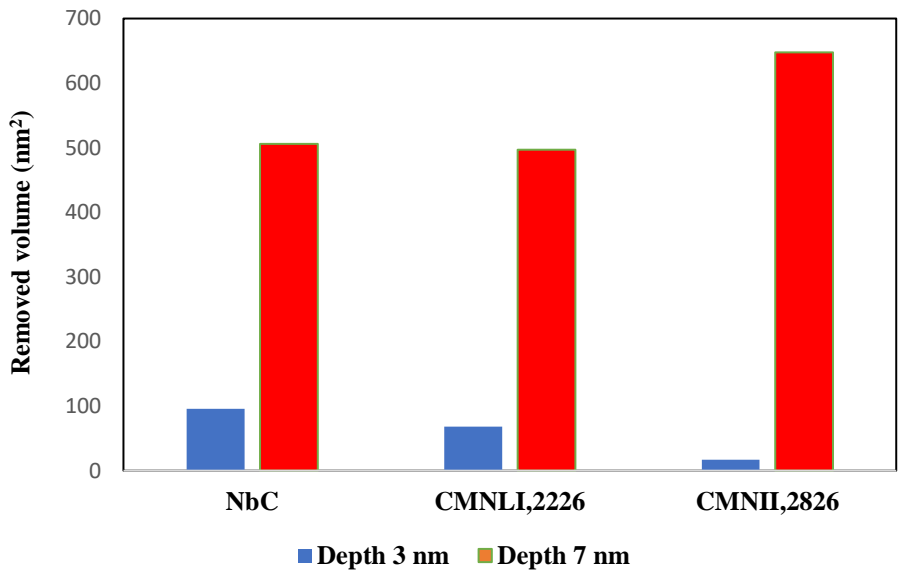

Figure 1: Comparison of material removed volume at different penetration depth

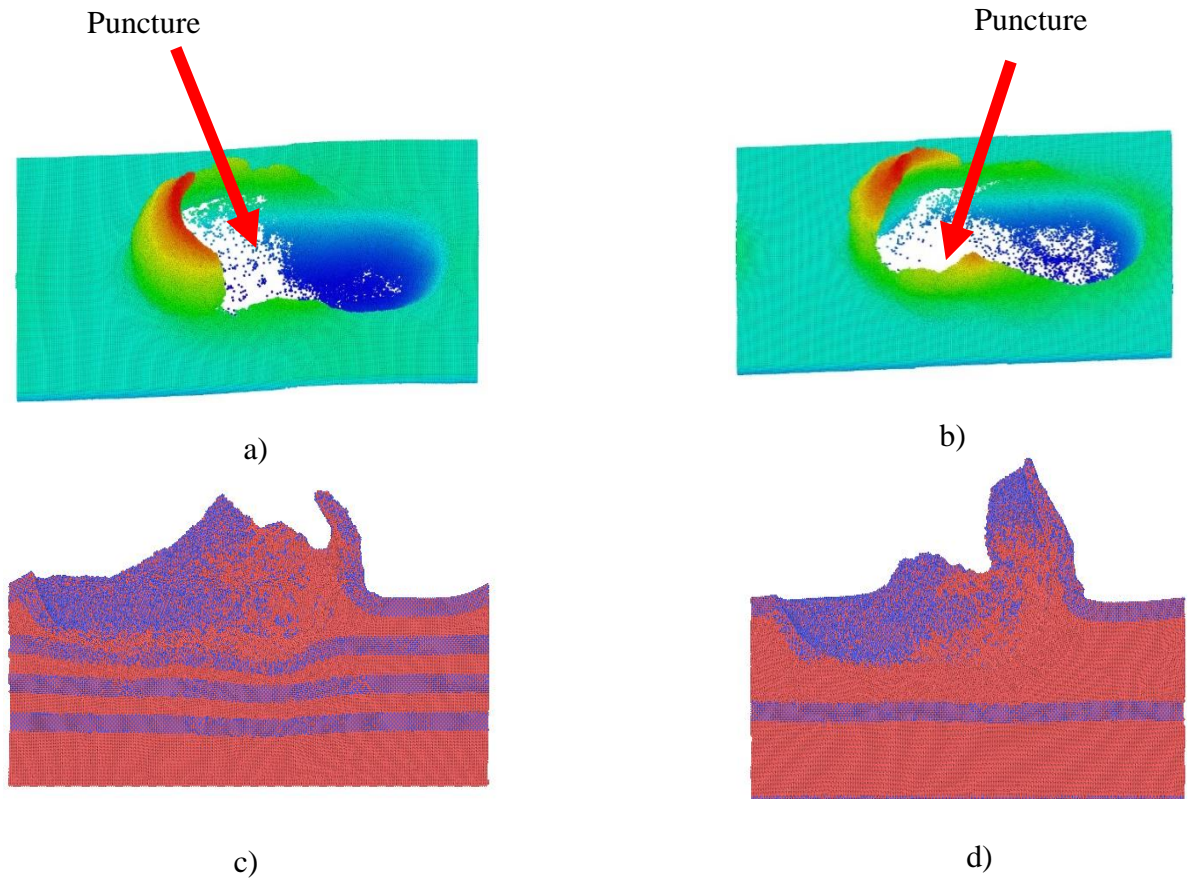

Figure 2: Perspective (a, c) and front view (b, d) of scratched surface at end of the scratching distance for CMNLI, 2226 (a, c) and CMNLII, 2826 (b, d). Atoms are colored based on the width for perspective view.

Figure 3 compares the scratching load, normal load, and friction coefficient at different penetration depths. It can be seen from Figure 15 that all tribological properties are significantly affected by the penetration depth. The scratching load and normal load of Model 22 become increasingly higher when the scratch is performed at deeper penetration depth, as shown in Figure 15 (a, b). However, the rate of increase in scratching load is more pronounced than the increasing rate of the normal load. This behavior is in good agreement with reported data [52] and it is linked to the fact that within the increase of penetration depth, both normal and transverse area increase but the increase of transverse contact area is more rapid. However, the normal load of CMNLII, 2826 at higher penetration depth is continuously decreasing and then it became saturated. This is mainly the indenter directly contacts with soft Nb metal due to the puncture at the top NbC (Figure 14), therefore, decreasing the normal force.

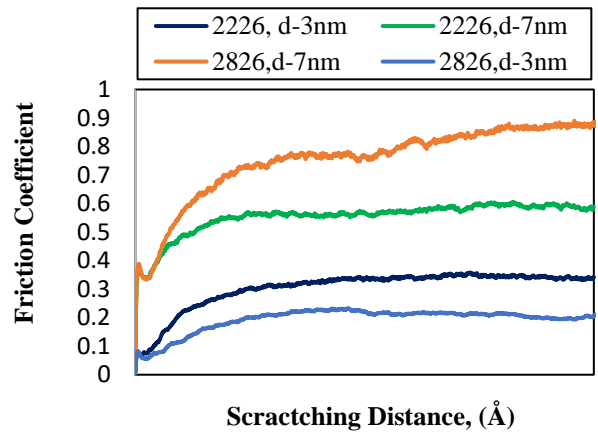

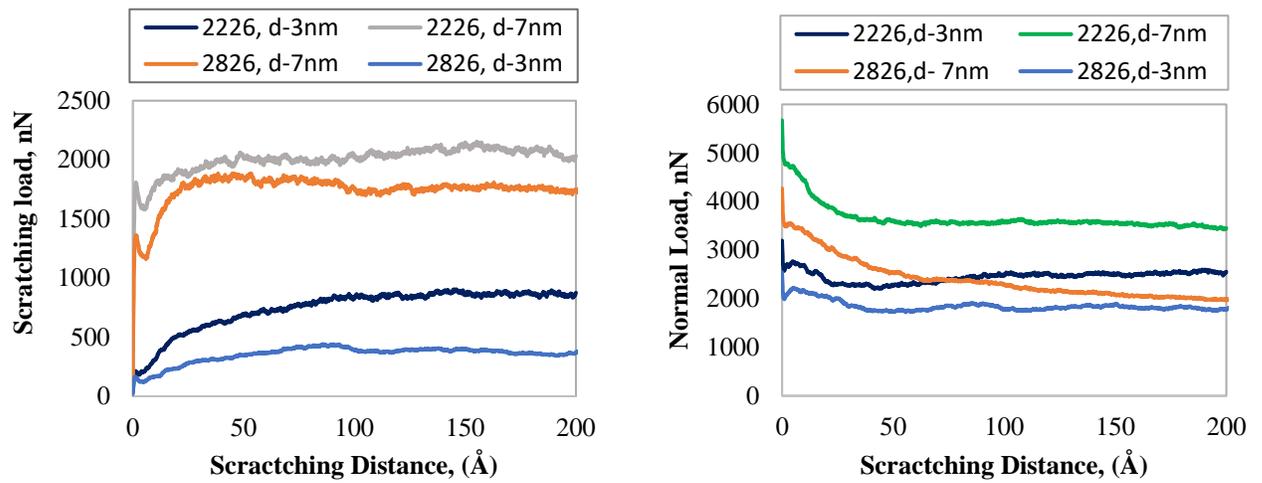

c)

Figure 3: The variations of a) scratching load, b) normal load and c) friction coefficient at different model

As pointed out in the introduction, the majority of the available literature has reported the scratching behavior of materials for extremely low penetration depth. To understand the scratching behaviors at high penetration depth, CMNL models were also compared with the single crystal Nb and NbC.

Fig. 4 shows the variation of the friction coefficient with the scratching distance for all the models including multilayers and single crystals at the 7nm penetration depth. The NbC single crystal exhibits the lowest friction coefficient. However, CMNLII model with the thickest metallic layer shows the highest friction properties, which also contradicts our previous study at 3nm penetration depth. To explain this behavior, one the scratching and normal load are observed. Fig. 5a and b show the variation of the scratching (friction) load and the normal load with the scratching distance, respectively, for all models at higher penetration depth. The scratching load and normal load for NbC single crystal are still the highest followed by CMNLI,2226, CMNLII,2826, and Nb single crystal which is good agreement with our previous study []. The variation in tangential force observed in CMNL (Ceramic Matrix Nanocomposite) models can be attributed to the Nb layer situated beneath the top ceramic layer, given that the top layer remains constant across all models. It has been noted that the C2/M2 models exhibit a thinner metallic layer which leads to a higher resistance against dislocation propagation, thus increasing the friction force compared to the C2/M8 model. This can be attributed to the greater difficulty in nucleating dislocations within the metallic layer of C2/M2 models. In contrast, the C2/M8 model showcases dislocations with greater freedom of movement within the metallic layer, owing to the greater distance between opposing interfaces, resulting in a lower friction force. These mechanisms contribute to the difference in friction force between the two models. However, it is noticeable that the normal load is very high compared to the scratching load for both models. Therefore, the normal load dominates the friction properties when the penetration depth is deeper, leading to a higher friction coefficient for the CMNLII model with the thickest metallic layer.

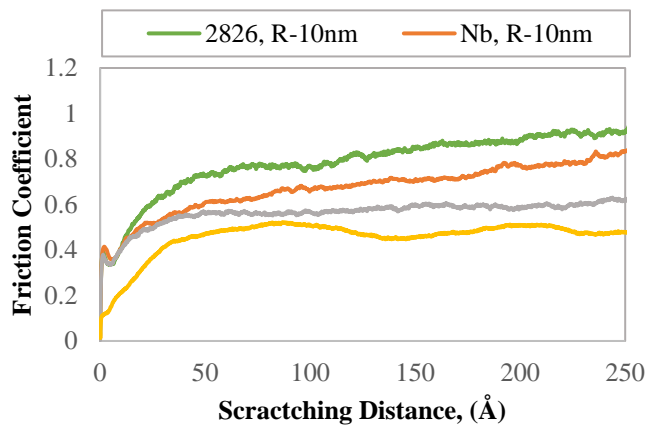

Figure 4: The variations of friction coefficient of different model at penetration depth 7 nm.

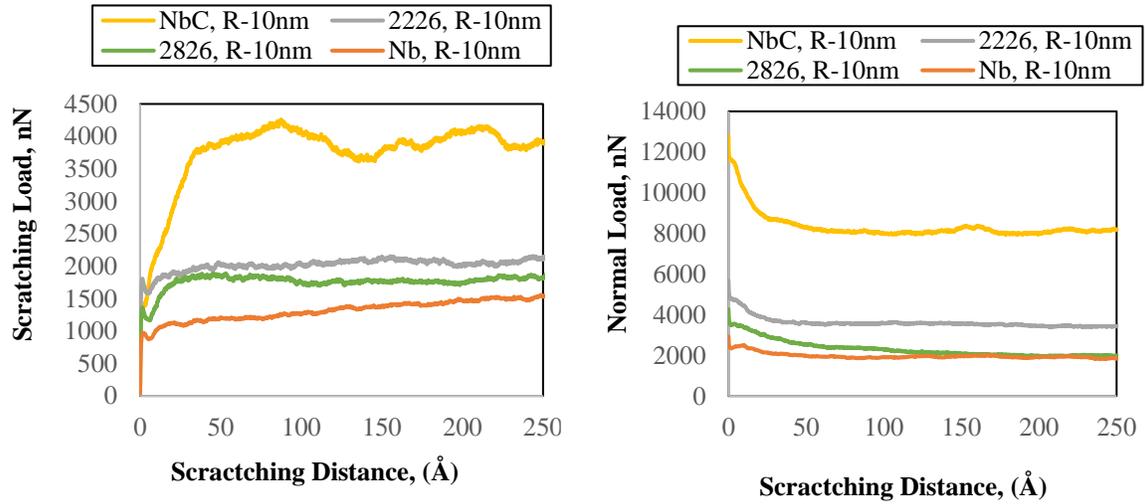

Figure 5: The variations of a) scratching load, and b) normal load of different model at penetration depth 7 nm.

To understand the effect of the metal-ceramic ratio on the scratching behavior more detailed, more simulations were carried out for multilayers having the same bi-layer thickness but with different individual layer thicknesses. Figure 6 shows the removed volume of material for each model. The contibution of individual ceramic and mettalic layers on total removed volume was also counted. It is shown that the removed volume are almost the same for all models except CMNL,2826 and CMNL,1926. That indicates the low dependency of pile-up atoms to the metal-ceramic thickness ratio at higher penetration depth. When the penetration depth is higher, all models produce a high amount of material removal, no matter the metal/ ceramic ratio. However, the CMNL,2826 and CMNL, 1926 show a very high amount of removed volume of material than the other models. Also, while only ceramic layer contributes to the total removed volume for other models, both metallic and ceramic layer contribute to the material removed volume for CMNL,1926, CMNL, 2826 models.

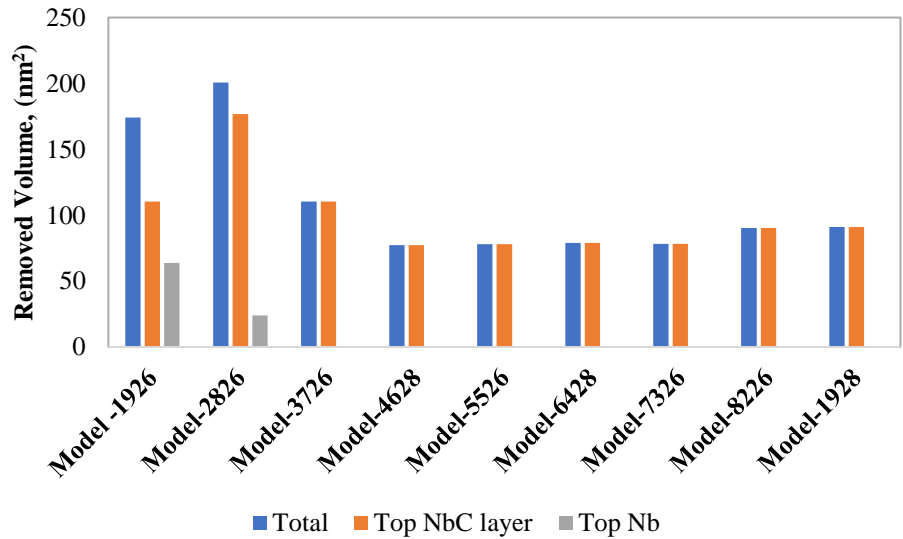

Fig. 6- Removed volume of the different model at the scratching length 20 nm

To understand this observed behavior, the atomic snapshot at the end of scratching was observed for each model. Figure 7 shows the perspective view of the top ceramic layer for CMNL,1926, CMNL, 2826 and CMNL,3726. For the sake of better comparison, only those models are kept which have a puncture during the scratching. Here, only CMNL,1926 and CMNL,2826 show the puncture at top ceramic layer during the scratching, but it is not seen in the other models. The reason for being punctured in only CMNL,1926 and CMNL,2826 models is the low thickness of the top ceramic layer. The thicknesses of the top ceramic layer in the CMNL,1926, CMNL, 2826 are too thin that the indenter can easily puncture the top ceramic during the indentation. Once the puncture initiates in the ceramic layer, it is very easy for both the top ceramic and metallic atoms to be piled up, creating a significant amount of pile-up atoms. Therefore, some contribution to the material removed volume from the metallic layer are seen for these two models as shown in Figure 18. With the increase of ceramic layer thickness, the number of ceramic atoms become enough to resist the indenter to puncture the top ceramic layer. The least thickness of the ceramic layer that can resist the indenter making a puncture can be defined as a critical thickness. For this specific study, the critical thickness is between 2 and 3 nm for indenter radius 10 nm at 7 nm penetration depth. This critical thickness depends on the indenter radius and penetration depth. For a constant indenter size, the critical thickness increases with the increase of penetration depth while it decreases with the indenter radius.

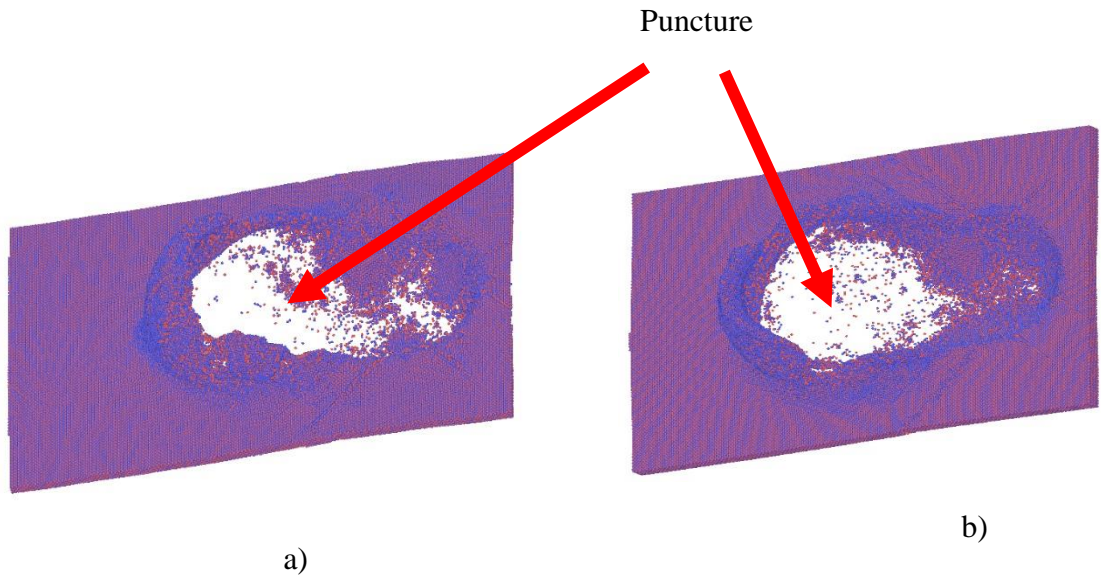

a)                                      b)

Puncture

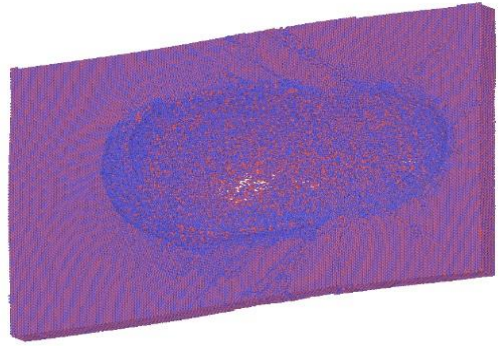

c)

Figure 7: Perspective view of scratched surface at end of the scratching distance showing puncture. a) CMNL,1926, b) CMNL,2826, c) CMNL, 3726.

Figure 8 shows the variation of scratching load, normal load, and friction coefficient with the scratching distance. The obtained results show similar behavior to our study at penetration depth 7 nm. The scratching load and normal load are still highest for CMNL models with the thinnest metallic layer as like our previous study. As the rate of variations of the normal load is more rapid, the normal load dominates the friction properties, leading to a higher friction coefficient for the model with the thickest metallic layer.

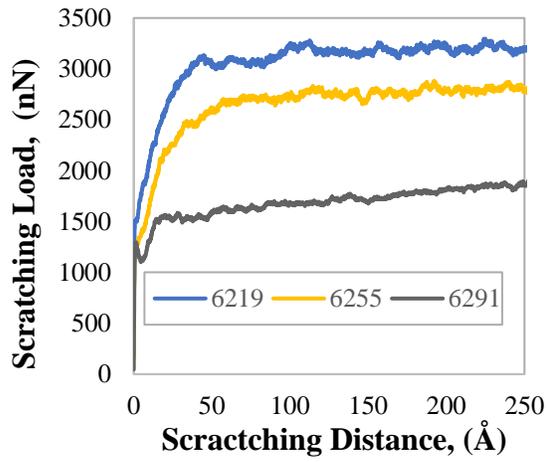
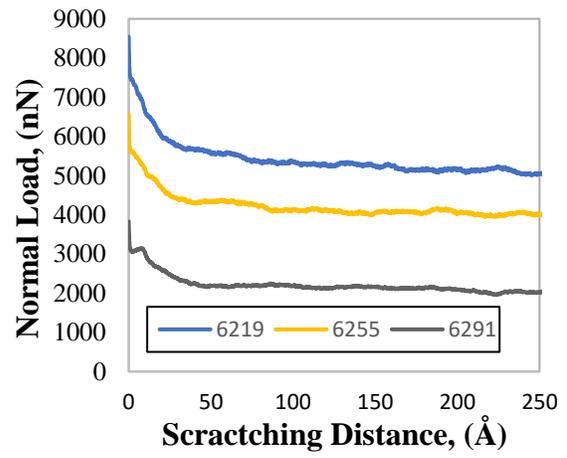

a)

b)

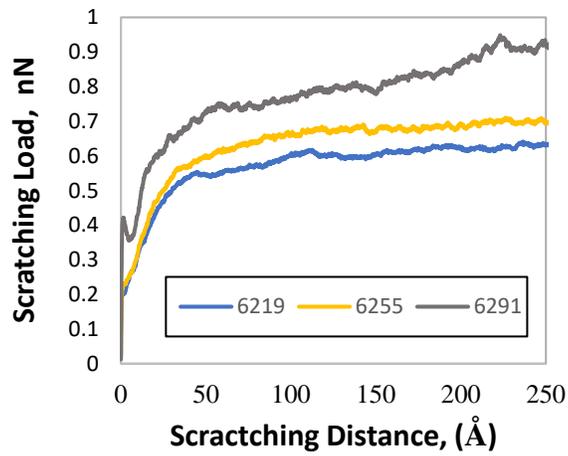

c)

Figure 8: The variations of a) scratching load, b) normal load and c) friction coefficient at different model

## 4. PREDICTION OF FRICTION COEFFICIENT USING MACHING LEARNING

Molecular dynamics simulation cost is very expensive and time consuming. Therefore, we aimed to predict the frictional coefficient using a machine-learning approach based on a dataset of 35 unique simulation models. If we could generate scratching load, normal load and friction coefficient by molecular dynamics simulation, that will allow us to reduce the computational cost and we can study any different model. By leveraging machine learning, we sought to enhance our understanding of material design and identify the proper material strength for various applications. The dataset was divided into training, validation, and testing sets, with 23 models (70%) used for training, five models (15%) for validation, and the remaining seven models (15%) for testing. We generated features from the simulated model's intended radius, depth, and geometry to build our predictive model. After training the model, we obtained a validation Mean Squared Error (MSE) of approximately 0.00098, indicating that the model performed well on the validation data. When evaluating the model on the test data, we found the following scores:

Mean Squared Error: 0.0030

Root Mean Squared Error: 0.055

Mean Absolute Error: 0.045

R-squared: 0.958

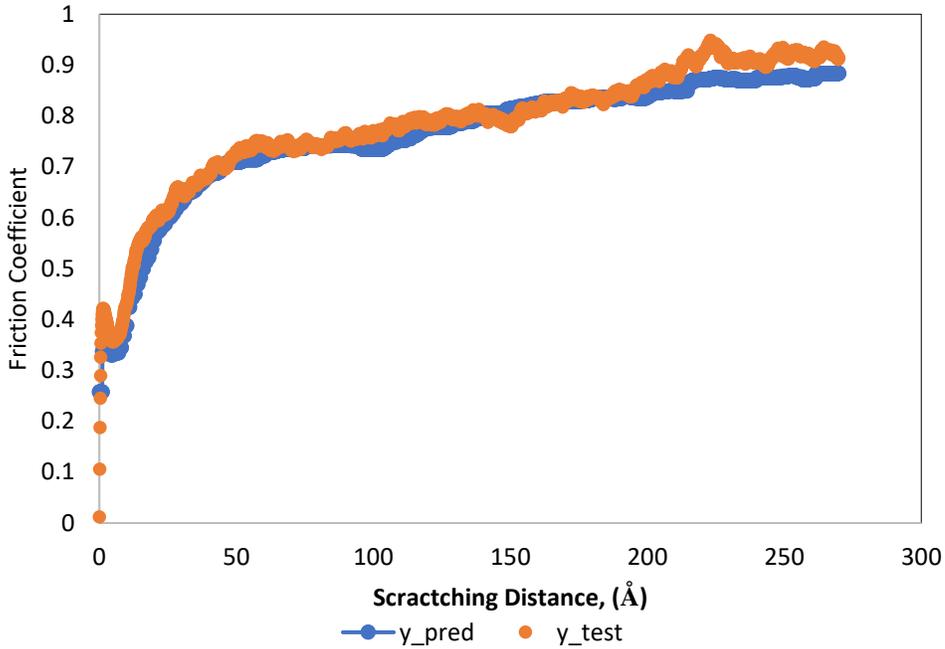

Figure 9: The comparison of friction coefficient (Molecular Dynamics Simulation vs Machine Learning)

Figure 9 suggests that our machine learning model could accurately predict the frictional coefficient, capturing the underlying relationships between the input features and the target variable. The model identified several important features that contributed to predicting the frictional coefficient. These features were primarily related to the simulation models' depth, radius, and geometry. By leveraging machine learning, we can optimize the material design and identify the proper material strength, leading to more efficient and effective systems in various applications. However, there are some limitations to our approach. Firstly, our dataset is relatively small, with only 35 simulation models. This may limit the model's ability to generalize to a broader range of scenarios. Secondly, although our model has identified important features, further investigation is needed to understand the underlying mechanisms that govern the relationship between these features and the frictional coefficient. Despite these limitations, our study demonstrates the potential of using machine learning to gain insights into material design and frictional behavior. Thus, we can reduce the computational cost of molecular dynamics simulation. In future study, we will improve our model and predict scratching, normal, friction coefficient and will validate different models data obtained from the molecular dynamics simulation.

## 5. CONCLUSION

Molecular dynamics atomistic simulations were performed to investigate the scratching behavior of NbC/Nb ceramic/metal nanolaminates with various metal thicknesses. Three models with different metal thicknesses were considered and their tribological behavior was compared to that of NbC and Nb single crystals. Nano-scratching was simulated by penetrating and moving a spherical indenter into the CMNLs. The results showed a significant effect of the metal thickness on the mechanical response of the models. The material removal rates of the CMNL models were lower than those for NbC and Nb single crystals proving the improving effect of the alternating metallic and ceramic layers on the scratch behavior of the hybrid materials. The scratching response of the models was linked to the underlying deformation mechanisms during the scratching process. The strain hardening of the metallic layers and their compliance was introduced as the significant factors affecting the scratching behavior of the models.

**ACKNOWLEDGMENT**

We would like to thank the Center for Research Computing & Data at Northern Illinois University for their support and providing us the access to their compunter cluster, GAEA.

**Figure Captions List**

Fig. 1- The simulation cell used for nanoindentation and scratching. The layers are numbered 1-4 from top to bottom. Only the thickness of layer 2 was varied.

Fig. 2- Load-displacement curves for CMNL models.